\pdfoutput=1
\RequirePackage{fix-cm}

\let\latexKernelLabel\label
\documentclass[%
 reprint,
 amsmath,amssymb,
 superscriptaddress,
 aps,
 preprintnumbers,
prl,
floatfix,
]{revtex4-2}

\DeclareRobustCommand{\timesbox}{%
  \mathbin{%
    \scalebox{0.8}{
    \ooalign{%
      $\Box$\cr
      \hfil\raisebox{0.2ex}{$\times$}\hfil\cr %
    }%
    }
  }%
}
\usepackage{booktabs}
\usepackage{multirow}
\usepackage{slashed}
\usepackage[normalem]{ulem}
\usepackage{graphicx}%
\usepackage{dcolumn}%
\usepackage{bm}%
\usepackage[nopatch=footnote]{microtype}
\usepackage{braket}
\usepackage{etoolbox}
\makeatletter
\IfFormatAtLeastTF{2023-06-01}{%
  \let\label\latexKernelLabel
  \let\ltx@label\label
}{}
\newcommand{\patchAPSBibControl}[2]{%
  \patchcmd{\@bibdataout@aps}{#1}{#2}{}{%
    \PackageError{ludy-paper}{REVTeX bibliography control patch failed}%
      {Check the installed REVTeX bibliography control definition.}%
  }%
}
\patchAPSBibControl{author="08"}{author="48"}
\patchAPSBibControl{author="08"}{author="48"}
\patchAPSBibControl{year="1"}{year="0"}
\patchAPSBibControl{year="1"}{year="0"}
\makeatother
\usepackage{hyperref}
\newcommand{\Eprint}[2]{%
  \begingroup
  \def\eprintText{#2}%
  \patchcmd{\eprintText}{arXiv:}{arXiv:\allowbreak}{}{}%
  \href{#1}{\eprintText}%
  \endgroup
}
\usepackage[table]{xcolor}

\usepackage{pgfplots}
\pgfplotsset{compat=1.18}
\usepackage{cleveref}
\usepackage{fancyvrb}
\usepackage{ragged2e}
\newcommand{\topblackrule}{%
  \noalign{\color{black}\hrule height 0.65pt}%
}
\newcommand{\midblackrule}{%
  \noalign{\color{black}\hrule height 0.40pt}%
}
\newcommand{\bottomblackrule}{%
  \noalign{\color{black}\hrule height 0.65pt}%
}

\definecolor{ludybg}{RGB}{232,243,247}   %
\definecolor{refbg}{RGB}{248,238,230}    %
\definecolor{mgfivebg}{RGB}{248,238,230} %
\usetikzlibrary{patterns}
\allowdisplaybreaks
\begin{document}

\title{The KLN theorem meets collinear factorisation: \texorpdfstring{\\}{ }Local Unitarity for hadronic cross-sections
}

\author{Zeno Capatti}
\affiliation{%
CERN, Theory Department, 1211 Geneva 23, Switzerland
}
\affiliation{%
  Institute for Theoretical Physics, University of Bern, 3012 Bern, Switzerland
}

\author{Lucien Huber}

\affiliation{%
  Institute for Theoretical Physics, University of Bern, 3012 Bern, Switzerland
}

\preprint{CERN-TH-2026-289}

\begin{abstract}
We present an initial-state implementation of the Kinoshita--Lee--Nauenberg (KLN) theorem that is compatible with collinear factorisation.
In this formulation, hadronic cross-sections are expressed as convolutions of parton distribution functions with partonic cross-sections obtained by summing over the number of incoming partons.
By virtue of the KLN theorem, the partonic cross-sections are infrared finite by construction, without relying on dimensional regularisation as an infrared regulator.
The Local Unitarity formalism provides a local realisation of this infrared finiteness, allowing partonic cross-sections to be evaluated directly in four dimensions by Monte Carlo integration.
We demonstrate the framework at next-to-leading order and leading power.
Using Local Unitarity, we compute the Drell--Yan and $t\bar{t}$ production cross-sections numerically; using Reverse Unitarity, we obtain the deep inelastic scattering and Drell--Yan cross-sections analytically. In all cases, a universal scheme conversion relates our results to those obtained in the \(\overline{\mathrm{MS}}\) parton model.

\end{abstract}

\maketitle

\section{Introduction}
\label{sec:introduction}

Producing meaningful perturbative predictions for hadronic collisions requires connecting the scattering of asymptotic quarks and gluons to that of hadrons. A consistent framework must ensure infrared (IR) finiteness, allow systematic resummation of large logarithms arising from soft and collinear radiation, and provide a universal description of the nonperturbative structure of hadrons.

In conventional calculations of hadronic cross-sections, infrared finiteness relies on two complementary mechanisms.
Final-state infrared singularities cancel upon summing over the number of degenerate final-state massless particles, as guaranteed by the Kinoshita--Lee--Nauenberg (KLN) theorem~\cite{Kinoshita:1962ur,PhysRev.133.B1549}. The number of initial-state partons is instead kept fixed, breaking KLN cancellations. Collinear factorisation~\cite{PhysRevD.8.3633,DAVIDPOLITZER1974129,Altarelli:1979ub,Collins:1981ta,Collins:1981tt,Collins:1983ju,Collins:1983pk,Collins:1985ue,Collins:1989gx,Bodwin:1984hc,Ellis:1978ty} provides a complementary mechanism: initial-state collinear singularities factorise universally and are absorbed by renormalising parton distribution functions (PDFs), while soft singularities associated with initial--initial and initial--final dipoles cancel separately between real and virtual contributions. This asymmetric treatment raises the question of whether the KLN theorem can also be implemented in the initial state while preserving collinear factorisation.

In this paper we provide an answer: we investigate an initial-state implementation of the KLN theorem in which partonic cross-sections are defined as sums over degenerate initial and final states of varying massless-parton multiplicity, ensuring KLN-driven cancellation of infrared singularities in both states. Within the Local Unitarity (LU) framework~\cite{Capatti:2020xjc,Capatti:2021bsm,Capatti:2022tit,Capatti:2025khs}, these cancellations can be realised locally, enabling direct Monte Carlo integration of partonic cross-sections in four dimensions. We demonstrate our framework at next-to-leading order (NLO), computing the deep inelastic scattering (DIS) and Drell--Yan (DY) cross-sections analytically using Reverse Unitarity~\cite{Anastasiou:2002qz,Anastasiou:2002yz,Anastasiou:2003yy,Anastasiou:2013mca,Anastasiou:2013srw}, while Drell--Yan and $t\bar{t}$ production cross-sections are computed numerically using Local Unitarity.

At leading power in a virtuality expansion unique to our framework, partonic cross-sections are IR-finite, evolve according to DGLAP equations~\cite{Altarelli:1977zs,Dokshitzer:1977sg,Gribov:1968ia,Gribov:1972ri} and turn into the $\overline{\text{MS}}$ counterparts by a universal redefinition of the PDFs. Hence, we reconcile for the first time an initial-state implementation of the KLN theorem with collinear factorisation.
In turn, this framework provides a new way to compute $\overline{\text{MS}}$ hadronic cross-sections, and may offer a systematic method to study subleading power effects.

\section{Simplest clustering criterion}
\label{sec:clustering}

To relate the scattering of many partons to that of hadrons, we use the \emph{simplest clustering criterion}, which arises naturally from an infrared-safe density matrix description of the initial state~\cite{Capatti:2025gqj}.
It gives an analytically tractable definition of jet clustering, allowing calculations to be performed with existing analytic techniques.

Consider $n$ massless partons and the $n!$-normalised phase-space measure ($d=4-2\epsilon$):
\begin{equation}
\mathrm{d}\Pi_n=\frac{1}{n!}\prod_{i=1}^{n} \frac{\mathrm{d}^d q_i}{(2\pi)^{d-1}} \delta(q_i^2)\Theta(q_i^0).
\end{equation}
The $n$ partons are clustered in a compound object with momentum $p$ by the following integration measure~\cite{Capatti:2025gqj}:
\begin{equation}
\label{eq:1_clustering}
\mathrm{d}\Pi_n^\mathbf{p}=\mathrm{d}\Pi_n\frac{\hat{\delta}(q_{1n}-p)\delta_{\mathfrak{a},\mathfrak{q}_{1n}}}{4\pi\langle\mathfrak{a}\rangle},
\end{equation}
with $q_{1n}=q_1+...+q_n$, $\mathfrak{q}_{1n}=\mathfrak{q}_1+...+\mathfrak{q}_n$ and $\hat{\delta}(x)=(2\pi)^d \delta(x)$.
$\mathbf{p}=(p,\mathfrak{a})$ is the tuple of the momentum $p$ of the cluster and its quark number $\mathfrak{a}\in\mathbb{Z}$, i.e. the conserved QCD
charge equal to three times the baryon number.
For the $i$-th parton, $\mathfrak{q}_i=1$ for a quark,
$\mathfrak{q}_i=-1$ for an antiquark, and $\mathfrak{q}_i=0$
for a gluon.
Hence, $\langle\mathfrak{q}\rangle=2N_c$ for $\mathfrak{q}= \pm 1$, and $\langle\mathfrak{q}\rangle=(d-2)(N_c^2-1)$ for $\mathfrak{q}=0$.
Eq.~\eqref{eq:1_clustering} imposes that the momenta and quark numbers of the $n$ initial-state partons sum to the cluster momentum $p$ and quark number $\mathfrak{a}$.

We cluster $n$ partons into two compound objects with momenta $p_1$ and $p_2$ by distributing them according to:
\begin{equation}
\label{eq:2_clustering}
\mathrm{d}\Pi_n^{\mathbf{p}_1\mathbf{p}_2}=\hspace{-0.4cm}\sum_{(S_1,S_2)\in J_n}\hspace{-0.35cm}\frac{\mathrm{d}\Pi_n (-1)^{|S_1\cap S_2|}}{4\pi} \prod_{j=1}^2 \frac{\hat{\delta}(q_{S_j}-p_j)\delta_{\mathfrak{a}_j,\mathfrak{q}_{S_j}}}{\langle\mathfrak{a}_j\rangle},
\end{equation}
with $q_{S_i}=\sum_{j\in S_i} q_j$ and $\mathfrak{q}_{S_i}=\sum_{j\in S_i} \mathfrak{q}_j$, and
\begin{equation}
J_n=\left\{(S_1,S_2)\big|\, S_i\subset[n], \, S_1\cup S_2=[n]\right\},
\end{equation}
with $[n]=\{1,...,n\}$, and $\mathbf{p}_i=(p_i,\mathfrak{a}_i)$.
Each initial-state parton belongs to one or both clusters.
The momenta and quark numbers of all partons belonging to the $i$-th cluster must sum up to the momentum $p_i$ and quark number $\mathfrak{a}_i$.
As $p_i^2$ becomes small, the momentum of a parton contained within the $i$-th cluster will be either a) collinear to $p_i$, if it is contained only in one cluster, or  b) soft, if it is contained in both clusters.

\section{Cross-sections with many initial-state partons}
\label{sec:formalism}

We consider QCD corrections to photon-mediated DIS, where the incoming electron with momentum $p_e$ emits a virtual photon with momentum $q$.
The partonic cross-section is the contraction of the leptonic tensor $L_{\mu\nu}(p_e,q)$ and the partonic tensor $W_{\mathfrak{a}}^{\mu\nu}(x,q^2;\Lambda^2)$ that depends on the Bjorken variable $x$, on $q^2$, and an infrared scale $\Lambda^2$.
Given the collections of all partonic combinations:
\begin{equation}
 \mathfrak{S}_n=\bigcup_{j=1}^n \{q,\bar{q},g\}^j,
\end{equation}
and $\tilde{\mathfrak{S}}_n=\mathfrak{S}_n\cup \{\emptyset\}$, we write the partonic tensor as a sum over initial-state multiplicities:
\begin{equation}
\label{eq:partonic_tensor_MP}
W_{\mathfrak{a}}^{\mu\nu}(x,q^2;\Lambda^2)= \sum_{\vec{\mathfrak{s}}\in \mathfrak{S}_\infty}\int_0^{\Lambda^2} \frac{\mathrm{d}p^2}{2\pi}W_{\mathfrak{a},\vec{\mathfrak{s}}}^{\mu\nu}(p,q),
\end{equation}
where $p=\frac{Q^2}{4xq_-}\eta_++xq_-\frac{p^2}{Q^2}\eta_-$, $q=q_-\eta_-+\frac{q^2}{4q_-}\eta_+$ in light-cone coordinates ($\eta_\pm^2=0$, $\eta_+\cdot \eta_-=2$) and $Q^2=-q^2$.
For fixed initial-state multiplicities:
\begin{equation}
\label{eq:partonic_tensor_MP_fixed_mult}
W_{\mathfrak{a},\vec{\mathfrak{s}}}^{\mu\nu}=\text{disc}_{1-x}\int \mathrm{d}\Pi_{|\vec{\mathfrak{s}}|}^\mathbf{p}\mathcal{A}(\vec{\mathfrak{s}} \, \gamma^\star_\mu(q)\rightarrow \vec{\mathfrak{s}} \, \gamma^\star_\nu(q)),
\end{equation}
where $\mathbf{p}=(p,\mathfrak{a})$. The discontinuity operator arises from the application of the optical theorem to DIS.
All partonic spins and colours are summed over, and $\mathcal{A}$ is the forward-scattering amplitude, including disconnected diagrams.
The parton-model partonic cross-section is the $\mathfrak{S}_1$ term of the sum in Eq.~\eqref{eq:partonic_tensor_MP}.
The hadronic tensor $W_{H}^{\mu\nu}$ is:
\begin{equation}
W_{H}^{\mu\nu}=\sum_{\mathfrak{q}\in\mathfrak{S}_1}\int_x^1 \frac{\mathrm{d}\xi}{\xi} f_{\mathfrak{q}}(\xi;\Lambda^2)W_{\mathfrak{q}}^{\mu\nu}(x/\xi,q^2;\Lambda^2).
\end{equation}

In DY, instead, two protons scatter and produce QCD radiation and a lepton pair with measured invariant mass $q^2$.
At leading order in the electromagnetic coupling, the leptonic phase space factorises multiplicatively.
Hence we may consider a final-state photon with mass $q^2$ instead.
In our extension, the partonic cross-section reads:
\begin{equation}
\frac{\mathrm{d}\sigma_{\mathfrak{a}\mathfrak{b}}}{\mathrm{d}q^2}(\hat{s};\Lambda^2)= \! \sum_{\substack{\vec{\mathfrak{s}}\in\mathfrak{S}_{\infty} \\ \vec{\mathfrak{s}}'\in\tilde{\mathfrak{S}}_{\infty} }}\left[\prod_{i=1}^2\int_0^{\Lambda^2}\frac{\mathrm{d}p_i^2}{2\pi}\right]\frac{\mathrm{d}\sigma^{\vec{\mathfrak{s}}\vec{\mathfrak{s}}'}_{\mathfrak{a}\mathfrak{b}}}{\mathrm{d}q^2}(p_1,p_2),
\end{equation}
where $p_1=x_1\frac{\sqrt{s}}{2}\eta_++\frac{p_1^2}{2x_1\sqrt{s}}\eta_-$, $p_2=x_2\frac{\sqrt{s}}{2}\eta_-+\frac{p_2^2}{2x_2\sqrt{s}}\eta_+$ and $\hat{s}=x_1 x_2 s$.
At fixed multiplicities,
\begin{equation}
\label{eq:DY_partonic_fixed}
\frac{\mathrm{d}\sigma^{\vec{\mathfrak{s}}\vec{\mathfrak{s}}'}_{\mathfrak{a}\mathfrak{b}}}{\mathrm{d}q^2}=\frac{1}{\hat{s}}\int \mathrm{d}\Pi_{|\vec{\mathfrak{s}}|}^{\mathbf{p}_1\mathbf{p}_2}\int \mathrm{d}\Pi_f |\mathcal{A}(\vec{\mathfrak{s}}  \rightarrow \vec{\mathfrak{s}}' \gamma^\star(k))|^2.
\end{equation}
Here, $\mathbf{p}_1=(p_1,\mathfrak{a})$, $\mathbf{p}_2=(p_2,\mathfrak{b})$ are the momenta and quark numbers of the two colliding clusters.
Letting $q_{\vec{\mathfrak{s}}}$ and $q_{\vec{\mathfrak{s}}'}$ be the sum of the partonic momenta in the initial and final state, final states are distributed according to:
\begin{equation}
\mathrm{d}\Pi_f= N_{ee} \, \mathrm{d}\Pi_{|\vec{\mathfrak{s}}'|} \frac{\mathrm{d}^d k}{(2\pi)^d}\hat{\delta}(q_{\vec{\mathfrak{s}}}-q_{\vec{\mathfrak{s}}'}-k)\delta^+_{q^2}(k),
\end{equation}
 where $N_{ee}=2\alpha/(3 q^2)$ is the remainder of the leptonic integration.
 The Dirac delta function implements the measurement of the invariant mass of the photon, and specifically reads
\begin{equation}
\delta^\pm_{m^2}(p)=2\pi  \delta\left (p^2-m^2\right)\Theta(\pm p^0).
\end{equation}
Spin and helicities are summed over.
The hadronic cross-section is the convolution
\begin{equation}
\hspace{-0.05cm}\frac{\mathrm{d}\sigma}{\mathrm{d}q^2}\!=\! \! \left[\prod_{i=1}^2\sum_{\mathfrak{q}_i \in\mathfrak{S}_1}\hspace{-0.075cm}\int_0^1 \mathrm{d}x_i f_{\mathfrak{q}_i}(x_i;\Lambda ^2)\right] \!  \frac{\mathrm{d}\sigma_{\mathfrak{q}_1\mathfrak{q}_2}}{\mathrm{d}q^2} \!\left(\hat{s};\Lambda^2\right)\! .
\end{equation}
The $t\bar{t}$ cross-section is obtained by simply swapping the final-state photon with a $t\bar{t}$ pair and changing the final-state phase-space measure $\mathrm{d}\Pi_f$ accordingly. In the following, we use $f=f^{(0)}+\tfrac{\alpha_s}{2\pi}f^{(1)}+\mathcal{O}(\alpha_s^2)$ for the coupling expansion of $f$.

\section{Diagrammatics and a generalised optical theorem}
\label{sec:diagrammatics}

The newly defined partonic cross-sections expand into sums of disconnected interference diagrams with many incoming legs.
This diagrammatic proliferation has long complicated initial-state implementations of the KLN theorem~\cite{Axelrod:1985yi,Frye:2018xjj,Lavelle_2006,Capatti:2025gqj,khalil2017completediagrammaticimplementationkinoshitaleenauenberg}.
We conjecture that, at leading power (LP) in $\Lambda^2$, the sum over all interference diagrams with more than one initial-state parton per hadron gives a \emph{“blown-up” integral representation of PDF counterterms in a certain scheme}.
Hence, the diagrammatic outburst may be systematically re-absorbed in a PDF redefinition.
The conjecture is backed by a calculation of a large gauge-invariant subset of the DIS diagrams~\cite{Capatti:2025gqj}, showing that the longitudinal structure function is insensitive to the multi-partonic contributions. It is also motivated by the following broader argument.

At NLO, the multi-partonic contributions to the cross-section, Eq.~\eqref{eq:partonic_tensor_MP_fixed_mult} ($|\vec{\mathfrak{s}}|>1$) and Eq.~\eqref{eq:DY_partonic_fixed} ($|\vec{\mathfrak{s}}|>2$), are integrals of tree-level amplitudes.
As $p_i^2\to 0$, the external partons become soft or collinear, and the tree-level amplitudes factorise at leading power into universal soft and collinear factors multiplying the leading-order cross-section, with \emph{no hard remainder}~\footnote{Due to the clustering, hard contributions can only arise from parton-model diagrams.}.
The corresponding cross-section contributions factorise as convolutions of the Born cross-sections with the collinear and soft factors.
Under the non-trivial assumption that soft double logarithms vanish, we obtain the conjectured statement.

Hence, the $\overline{\text{MS}}$ result amounts to absorbing in the PDFs both the multi-partonic contributions and the $\overline{\text{MS}}$ poles of the bare parton-model cross-section.
We instead define PDFs $f_{\mathfrak{q}}^{\text{LU}}$ that absorb all those interference diagrams that are not strictly required for IR finiteness of the bare parton model.
The remaining calculation consists of the bare parton model together with the \emph{minimal} set of multi-partonic contributions needed for IR finiteness.

\begin{figure*}[t]
  \centering
   \vspace{-0.5cm}
  \includegraphics{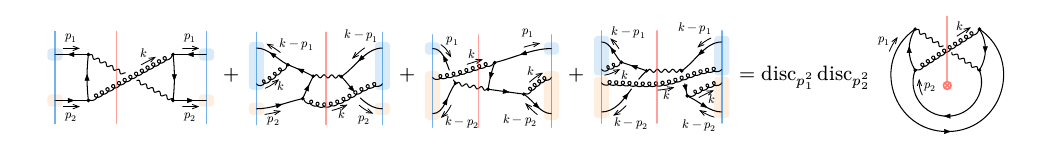}
  \vspace{-0.5cm}
  \caption{ \justifying Contribution of the $G=\timesbox$ cross-box diagram in the sum over $G\in \Gamma_{\mathfrak{q}\mathfrak{q}'}$ in \eqref{eq:disc_dy}, written as an (infrared-finite) discontinuity and as a sum of interference diagrams forming an infrared-finite class. The lepton phase space factorises and the photon is massive. The first diagram is the parton model's, while the other ones form the minimal set of multi-partonic diagrams needed for its IR finiteness. Coloured shadings denote belonging to the $p_1$ (blue) jet or the $p_2$ (orange) jet, as per the simplest-clustering criterion. The red line identifies the final-state cut. All initial- and final-state particles receive cut Feynman rules. On the right-hand side, we represent the integral $I_{\hspace{-0.075cm}\timesbox}^{\mathfrak{q}\mathfrak{q}'}$ (given in the \hyperref[sec:integrals_DIS_DY]{End Matter}) as a vacuum graph embedding~\cite{Capatti:2025gqj}.
  }
  \label{fig:xbox-sum}
  {\color{gray!50}\rule{0.8\linewidth}{0.6pt}}
  \phantom{g}\\[-0.3cm]
\end{figure*}

In the LU scheme, the contributing interference diagrams are packaged compactly (see, e.g., Fig.~\ref{fig:xbox-sum}).
Let $\Gamma_{\mathfrak{q}}$ denote the set of $\overline{\text{MS}}$ parton-model diagrams for DIS, and allow the initial-state partons to carry arbitrary virtualities.
At NLO and LP, the DIS partonic tensors read:
\begin{align}
\label{eq:disc_dis}
\left(W_{\mathfrak{q}}^{\text{LU},(1)}\right)^{\mu\nu}&=\sum_{G\in\Gamma_{\mathfrak{q}}}\int_0^{\Lambda^2} \frac{\mathrm{d}p^2}{8\pi^2 i}\frac{\text{disc}_{p^2}\text{disc}_{1-x}I_{\mathfrak{q} G}^{\mu\nu}}{\langle\mathfrak{q}\rangle}.
\end{align}
 The integrals $I_{\mathfrak{q} G}^{\mu\nu}$ are given as \hyperref[sec:integrals_DIS_DY]{End Matter} material.
 The discontinuities in $p^2$ and $1-x$ account for the sums over initial- and final-state cuts, respectively.
 The operator $\text{disc}_{X}$ is both understood as a sum over Cutkosky cuts~\cite{Cutkosky:1960sp} in the $X$ channel (see Fig.~\ref{fig:xbox-sum}) and as the difference of the function it operates on evaluated at $X-i\varepsilon$ and $X+i\varepsilon$ (at LP, no subtleties such as those of Refs.~\cite{Bourjaily:2020wvq,Hannesdottir:2022xki,Abreu:2014cla,Abreu:2017enx} arise).

For hadronic collisions, the situation is analogous.
For DY, the NLO, LP cross-section reads:
\begin{align}
\label{eq:disc_dy}
\frac{\mathrm{d}\sigma_{\mathfrak{q}\mathfrak{q}'}^{\text{LU},(1)}}{\mathrm{d}q^2}&= \hspace{-0.2cm}\sum_{G\in\Gamma_{\mathfrak{q}\mathfrak{q}'}} \hspace{-0.15cm} \text{Re}\hspace{-0.05cm}\left[\prod_{i=1}^2\int_0^{\Lambda^2}\hspace{-0.1cm}\frac{\mathrm{d}p_i^2}{2\pi i}\text{disc}_{p_i^2}\right] \! \frac{N_{ee}I_{ G}^{\mathfrak{q}\mathfrak{q}'}}{4\pi \hat{s}\langle\mathfrak{q}\rangle \langle\mathfrak{q}'\rangle},
\end{align}
where the set $\Gamma_{\mathfrak{q}\mathfrak{q}'}$ is given in the \hyperref[sec:integrals_DIS_DY]{End Matter}.
There is one discontinuity for each incoming hadron. Since the discontinuity of a finite integral is still finite, we obtain that each individual graph summand is IR finite.

We determine the universal scheme transformation of PDFs from the LU scheme to the $\overline{\text{MS}}$ scheme by standard Altarelli--Parisi (AP) techniques~\cite{Altarelli:1977zs}, adapted to our framework (see \hyperref[sec:reproducible-analytic-calculations]{End Matter}).
This also proves the factorisability of the minimal set.
We introduce the \emph{splitting counterterms} $D_{\mathfrak{q}\mathfrak{q}'}$ through the convolution $f_{\mathfrak{q}'}^{\overline{\text{MS}}}=D_{\mathfrak{q}'\mathfrak{q}}\otimes f_{\mathfrak{q}}^{\text{LU}}$,
\begin{equation}
f_{\mathfrak{q}'}^{\overline{\text{MS}}}(x;\mu^2)=\hspace{-0.15cm}\sum_{\mathfrak{q}\in\mathfrak{S}_1}\hspace{-0.05cm}\int_x^1 \hspace{-0.05cm}\frac{\mathrm{d}\xi}{\xi} D_{\mathfrak{q}'\mathfrak{q}}\hspace{-0.06cm}\left(\xi;\tfrac{\Lambda^2}{\mu^2}\right)f_{\mathfrak{q}}^{\text{LU}}\hspace{-0.06cm}\left(\tfrac{x}{\xi};\Lambda^2\right).
\end{equation}
Their coupling expansion reads
\begin{equation}
D_{\mathfrak{q}\mathfrak{q}'}\left(\xi;\tfrac{\Lambda^2}{\mu^2}\right)=\sum_{n=0}^\infty \left(\frac{\alpha_s}{2\pi}\right)^{n} D_{\mathfrak{q}\mathfrak{q}'}^{(n)}\left(\xi;\tfrac{\Lambda^2}{\mu^2}\right),
\end{equation}
with $D_{\mathfrak{q}\mathfrak{q}'}^{(0)}=\delta_{\mathfrak{q}\mathfrak{q}'}\delta(1-\xi)$, and:
\begin{widetext}
\begin{align}
D_{\mathfrak{q}\mathfrak{q}'}^{(1)}(\xi)=
\left.
\sum_{\mathfrak{a}\in\mathfrak{S}_1}\int_0^1 \mathrm{d}\xi'\,
\frac{c_{\mathfrak{a}\mathfrak{q}'}}{\Gamma(1-\epsilon)}
\Bigg\{\left[\delta_{\mathfrak{q}\mathfrak{a}}\delta(\xi-\xi')
-\frac{\delta_{\mathfrak{q}\mathfrak{q}'}\delta(1-\xi)}2\right]
\frac{P_{\mathfrak{a}\mathfrak{q}'}^{(0)}(\xi')}{\epsilon}-\delta_{\mathfrak{q}\mathfrak{a}}\delta(\xi-\xi')b_\partial^{\mathfrak{a}\mathfrak{q}'}
\Bigg\}
\left(\frac{\Lambda^2\xi'\bar{\xi}'}
{e^{\gamma_E}\mu^2}\right)^{-\epsilon}
\right|_{\epsilon^0}. \label{eq:splitting_cts_nlo}
\end{align}

\end{widetext}
Here, $\bar{\xi}'=1-\xi'$, $c_{qg}=T_F$, $c_{qq}=c_{gq}=C_F$ and $c_{gg}=C_A$.
 The function $P_{\mathfrak{q}\mathfrak{q}'}^{(0)}(\xi)$ is the leading-order AP splitting kernel in conventional dimensional regularisation, Eqs.~(32)--(35) of Ref.~\cite{Catani:1996pk}, stripped of colour factors, while $\big|_{\epsilon^0}$ stresses that the pole is discarded.
 The $\delta(\xi-\xi')$ pieces originate from the contribution with an additional QCD emission in the final state.
 Finally, $b_{\partial}^{qg}=1$, $b_{\partial}^{gq}=b_{\partial}^{qq}=1-\epsilon$, $b_{\partial}^{gg}=\tfrac{3(2\epsilon-3)}{4(1-\epsilon)}$ specify the Feynman-gauge boundary terms.
 Reverse Unitarity calculations need them, whereas Local Unitarity uses $b_{\partial}^{\mathfrak{q}\mathfrak{q}'}=0$.

\section{Reverse Unitarity computation}
\label{sec:reverse-unitarity}

We now compute the cut diagrams for the NLO DIS and DY cross-sections in the LU scheme using Reverse Unitarity, providing two benchmark tests of the correctness and universality of the splitting counterterms of Eq.~\eqref{eq:splitting_cts_nlo}.

\underline{\textbf{DIS}}: The structure functions are obtained from the partonic tensor by contraction with suitable projectors, $F_{a\mathfrak{q}}=\mathrm{P}_a^{\mu\nu}(W_{\mathfrak{q}})_{\mu\nu}$.
We obtained in Ref.~\cite{Capatti:2025gqj} that, for a quark:
\begin{align}
\frac{F_{2q}^{\text{LU},(1)}}{e_q^2C_F x}
&=
\tilde{P}_{qq}^{(0)}(x)L_x-\frac{2\pi^2+3}{3}\delta(1-x)
\nonumber\\
&\hspace{1cm}
-\frac{6x^2-4x+1}{2(1-x)_+}
+\mathcal{O}\left(\tfrac{\Lambda^2}{q^2_{\text{h}}}\right),
\\
\frac{F_{Lq}^{\text{LU},(1)}}{e_q^2C_F x}
&= 2x+\mathcal{O}\left(\tfrac{\Lambda^2}{q^2_{\text{h}}}\right).
\end{align}
For a gluon:
\begin{align}
\frac{F_{2g}^{\text{LU},(1)}}{2e_q^2 T_F x}&= \tilde{P}_{qg}^{(0)}(x)L_x+6x(1-x)-2+\mathcal{O}\left(\tfrac{\Lambda^2}{q^2_{\text{h}}}\right), \\
\frac{F_{Lg}^{\text{LU},(1)}}{2e_q^2 T_F x}&=4  x(1-x)+\mathcal{O}\left(\tfrac{\Lambda^2}{q^2_{\text{h}}}\right),
\end{align}
where $L_x=\log\left(\tfrac{Q^2}{\Lambda^2 x^2}\right)$ and $q_{\mathrm{h}}^2=\min(Q^2,(p+q)^2)$.
$\tilde{P}^{(0)}_{\mathfrak{q}\mathfrak{q}'}$ is the regularised Altarelli--Parisi splitting kernel, Eqs.~(C.6)--(C.9) of Ref.~\cite{Catani_1997}, stripped of colour factors.

The result is IR finite, owing to the KLN mechanism, and the scale evolution and longitudinal structure functions take the same form as in the $\overline{\text{MS}}$ parton model.
Hence, structure functions can be cast in the $\overline{\text{MS}}$ scheme by a scheme redefinition.
More precisely, we find that
\begin{equation}
F_{2\mathfrak{q}}^{\overline{\text{MS}}}=\sum_{\mathfrak{q}'\in\mathfrak{S}_1}D^{-1}_{\mathfrak{q}'\mathfrak{q}}\otimes F_{2\mathfrak{q}'}^{\text{LU}}+\mathcal{O}\left(\alpha_s^2,\tfrac{\Lambda^2}{q^2_{\text{h}}}\right),
\end{equation}
where the $D_{\mathfrak{q} \mathfrak{q}'}$ functions are precisely those of Eq.~\eqref{eq:splitting_cts_nlo}, with the Feynman-gauge boundary terms, and the $\overline{\text{MS}}$ result is given in Ref.~\cite{Bardeen:1978yd}.

\underline{\textbf{DY}}: We can compute the NLO DY cross-section in the same formalism.
The Reverse Unitarity computation gives, using $q^2=z \hat{s}$,
\begin{align}
&\frac{\mathrm{d}\sigma_{q\bar{q}}^{\text{LU},(1)}}{\mathrm{d}q^2}= \frac{e_q^2 \alpha C_F}{3N_cq^2 \hat{s}}\! \Bigg[ 2\tilde{P}_{qq}^{(0)}(z)L_z -\delta(1-z)\!\nonumber \\
&+2(1+z^2)\! \left[\frac{\log(1\!-\! z)}{1\!-\!z}\right]_+ +2z-4 \Bigg] \!  +\mathcal{O}\left(\tfrac{\Lambda^2}{q^2_{\text{h}}}\right), \label{eq:qq_DY}\\
&\frac{\mathrm{d}\sigma_{gq}^{\text{LU},(1)}}{\mathrm{d}q^2}=\frac{e_q^2 \alpha T_F}{3N_cq^2 \hat{s}}\Bigg[
\tilde{P}_{qg}^{(0)}(z)\left(L_z+\log(1-z)\right) \nonumber
 \\
 &+\frac{-1+2z-3z^2}{2}\Bigg] +\mathcal{O}\left(\tfrac{\Lambda^2}{q^2_{\text{h}}}\right), \label{eq:qg_DY}
\end{align}
where $L_z=\log\left(\tfrac{\hat{s}}{\Lambda^2z}\right)$ and $q_{\mathrm{h}}^2=\min(q^2,\hat{s}-q^2)$.
Equations~\eqref{eq:qq_DY} and \eqref{eq:qg_DY} are related to the $\overline{\text{MS}}$ results~\cite{Altarelli:1979ub} by
\begin{align}
\hspace{-0.05cm}\frac{\mathrm{d}\sigma_{\mathfrak{q}\mathfrak{q}'}^{\overline{\text{MS}}}}{\mathrm{d}q^2}=\hspace{-0.3cm}\sum_{\mathfrak{q}_1,\mathfrak{q}_2\in\mathfrak{S}_1}\hspace{-0.3cm}D^{-1}_{\mathfrak{q}_1\mathfrak{q}}\otimes\frac{\mathrm{d}\sigma_{\mathfrak{q}_1\mathfrak{q}_2}^{\text{LU}}}{\mathrm{d}q^2}\otimes D^{-1}_{\mathfrak{q}_2\mathfrak{q}'}\hspace{-0.05cm}+\mathcal{O}\left(\alpha_s^2,\tfrac{\Lambda^2}{q^2_{\text{h}}}\right),
\end{align}
which uses $D_{qq}$ and $D_{qg}$ as defined by Eq.~\eqref{eq:splitting_cts_nlo}.

These computations confirm that, at NLO, LU and $\overline{\text{MS}}$ cross-sections carry the same scale dependence and are related to each other by a universal scheme redefinition.

\section{Local Unitarity computation}
\label{sec:local-unitarity}

\begin{figure}[!tp]
\begin{minipage}{\columnwidth}
\begin{tikzpicture}
\begin{axis}[
  name=top,
  width=0.74\linewidth, height=88pt,
  scale only axis=true,
  xmin=0, xmax=1, ymin=-7.6, ymax=8.3,
  xtick=\empty,
  ytick={-7,-4,0,2,4,6,8},
  yticklabels={$-10^{-6}$,,$0$,,$10^{-9}$,,$10^{-5}$},
  ylabel={$\Delta\sigma^{(1)}\,[\mathrm{pb}]$},
  label style={font=\fontsize{7.4}{8.2}\selectfont},
  tick label style={font=\fontsize{6.95}{7.5}\selectfont},
  yticklabel pos=right,
  ylabel style={at={(axis description cs:-0.02,0.5)},anchor=south},
  tick align=inside, tick style={line width=0.45pt},
  axis line style={line width=0.5pt},
  legend style={draw=none, fill=none, font=\fontsize{7.2}{8.0}\selectfont,
    at={(0.02,0.98)}, anchor=north west, cells={anchor=west},
    clip=false,
    inner sep=0pt, row sep=1.0pt},
  legend image post style={xscale=0.82},
  scaled ticks=false,
]
\addplot[forget plot, densely dotted, line width=0.35pt] coordinates {(0,0) (1,0)};
\addplot[forget plot, const plot, solid, line width=0.65pt, no markers] coordinates {
  (0,1.05489884258e-06) (0.1,1.81655782870) (0.2,3.69055227540) (0.3,4.68686208211)
  (0.4,5.30341026097) (0.5,5.74994035069) (0.6,5.99411306404) (0.7,5.68622316265)
  (0.8,-6.97754061265) (0.9,7.82037785875) (1,7.82037785875)
};
\addlegendimage{solid,line width=0.65pt,mark=*,mark size=1.35pt}
\addlegendentry{\texttt{LU}}
\addplot[forget plot, only marks, mark=*, mark size=1.35pt,
  mark options={solid,line width=0.5pt},
  error bars/.cd, y dir=both, y explicit,
  error bar style={solid,line width=0.5pt}, error mark options={rotate=90,mark size=1.4pt,line width=0.5pt}
] table[x=x,y=y,y error plus=ep,y error minus=em,row sep=\\] {
x y ep em\\
0.041 1.05489884258e-06 2.64087947206e-08 2.64087947206e-08\\
0.141 1.81655782870 0.00273902912079 0.00275641345844\\
0.241 3.69055227540 0.00271393559600 0.00273100186730\\
0.341 4.68686208211 0.0129164230603 0.0133123787696\\
0.441 5.30341026097 0.00677970475024 0.00688722234322\\
0.541 5.74994035069 0.00555511370461 0.00562709151277\\
0.641 5.99411306404 0.00535913117347 0.00542608919055\\
0.741 5.68622316265 0.0219867757092 0.0231595233173\\
0.841 -6.97754061265 0.00228931044430 0.00227730598052\\
0.941 7.82037785875 0.000378038121699 0.000378367477344\\
};
\addplot[forget plot, const plot, densely dashed, line width=0.65pt, no markers] coordinates {
  (0,1.04029000540e-06) (0.1,1.81770713663) (0.2,3.69115121909) (0.3,4.68008690488)
  (0.4,5.30825093191) (0.5,5.73984515445) (0.6,6.00123251685) (0.7,5.67704925620)
  (0.8,-6.97522112019) (0.9,7.82002671352) (1,7.82002671352)
};
\addlegendimage{densely dashed,line width=0.65pt,mark=triangle*,mark size=1.35pt}
\addlegendentry{\texttt{MG5aMC}}
\addplot[forget plot, only marks, mark=triangle*, mark size=1.35pt,
  mark options={solid,line width=0.5pt},
  error bars/.cd, y dir=both, y explicit,
  error bar style={solid,line width=0.5pt}, error mark options={rotate=90,mark size=1.4pt,line width=0.5pt}
] table[x=x,y=y,y error plus=ep,y error minus=em,row sep=\\] {
x y ep em\\
0.059 1.04029000540e-06 2.10132805804e-07 2.10132805804e-07\\
0.159 1.81770713663 0.00174164029503 0.00174865288338\\
0.259 3.69115121909 0.000465628351284 0.000466128110072\\
0.359 4.68008690488 0.000246551418797 0.000246691466934\\
0.459 5.30825093191 0.000196477658026 0.000196566586034\\
0.559 5.73984515445 0.000194158635128 0.000194245475828\\
0.659 6.00123251685 0.000256211641687 0.000256362882744\\
0.759 5.67704925620 0.00123209676085 0.00123560217602\\
0.859 -6.97522112019 0.000157543224345 0.000157486095212\\
0.959 7.82002671352 7.17258279925e-05 7.17376758157e-05\\
};

\node[
  anchor=south,
  align=center,
  font=\fontsize{7.8}{8.8}\selectfont\bfseries
] at (axis description cs:0.5,1.025)
{A - NLO correction to the hadronic DY cross-section \\ differential in $z$
($\sqrt{s}=1000\,\mathrm{GeV}$, $\sqrt{q^2}>300\,\mathrm{GeV}$)};

\end{axis}

\begin{axis}[
  name=pulls,
  at={(top.south west)}, anchor=north west, yshift=-3pt,
  width=0.74\linewidth, height=30pt,
  scale only axis=true,
  xmin=0, xmax=1, ymin=-3.0, ymax=3.0,
  xtick={0,0.2,0.4,0.6,0.8,1},
  xticklabels={$0$,$0.2$,$0.4$,$0.6$,$0.8$,$1$},
  ytick={-2,0,2},
  xlabel={$z$}, ylabel={pull},
  label style={font=\fontsize{7.05}{7.7}\selectfont},
  tick label style={font=\fontsize{6.7}{7.25}\selectfont},
  xlabel style={at={(axis description cs:0.5,-0.25)},anchor=north},
  ylabel style={at={(axis description cs:-0.02,0.5)},anchor=south},
  tick align=inside, tick style={line width=0.45pt},
  yticklabel pos=right,
  axis line style={line width=0.5pt},
  clip=true,
]
\addplot[forget plot, densely dotted, line width=0.35pt] coordinates {(0,0) (1,0)};
\addplot[forget plot, densely dashed, line width=0.35pt] coordinates {(0,2) (1,2)};
\addplot[forget plot, densely dashed, line width=0.35pt] coordinates {(0,-2) (1,-2)};
\addplot[solid, line width=0.5pt, mark=square*, mark size=1.1pt] coordinates {
  (0.05,0.068979307967) (0.15,-0.353284099910) (0.25,-0.216991949962) (0.35,0.512680183335) (0.45,-0.712097570330)
  (0.55,1.78383020125) (0.65,-1.32962208778) (0.75,0.401931416367) (0.85,-1.01077190373) (0.95,0.911852516279)
};
\end{axis}

\node[
  anchor=north west,
  inner sep=-6pt
] at ([yshift=-30pt]pulls.south west) {%
  \begingroup
  \fontsize{6.9}{7.6}\selectfont
  \renewcommand{\arraystretch}{1.15}
  \setlength{\tabcolsep}{2.5pt}
  \begin{tabular*}{0.8\linewidth}{
    |@{\extracolsep{\fill}}c|c|c|
  }
  \hline
  \rule{0pt}{2.3ex}
  total $[10^{-6}\,\mathrm{pb}]$
  & \texttt{LU}: $5.8920(61)$
  & \texttt{MG5aMC}: $5.8912(12)$
  \\[1pt]
  \hline
  \end{tabular*}
  \endgroup
};

\end{tikzpicture}
\end{minipage}
\phantom{g}\\[0.1cm]
\begin{minipage}{0.45\textwidth}
\begin{minipage}{1\textwidth}
{\fontsize{7.8}{8.8}\selectfont\bfseries
B - NLO corrections to total $\mathfrak{q}\mathfrak{q}'\rightarrow t \bar{t}$ cross-section
\par}
\phantom{}\\[-0.2cm]
\begin{tabular}{
@{\hspace{5pt}}
l@{\hspace{10pt}}
l@{\hspace{10pt}}
>{\columncolor{ludybg}[4pt][4pt]}l@{\hspace{10pt}}
>{\columncolor{refbg}[4pt][4pt]}l@{\hspace{8pt}}
r@{\hspace{5pt}}
}
\topblackrule
\scalebox{0.9}{$\sqrt{\hat s}$ [$\scalebox{0.8}{$\text{GeV}$}$]}
& \scalebox{0.9}{$\mathfrak{q}\mathfrak{q}'$}
& \scalebox{0.9}{\texttt{LU} [$\scalebox{0.8}{$10^{-9}\,\text{GeV}^{-2}$}$]}
& \scalebox{0.9}{\cite{Czakon:2008ii} [$\scalebox{0.8}{$10^{-9}\,\text{GeV}^{-2}$}$]}
& \scalebox{0.9}{pull} \\
\midblackrule

\multirow{3}{*}{$400$}
 & $q\bar q$ & $-8.650(27)$
 & $-8.694003$ & $+1.65$ \\
 & $qg$ & $-0.40523(58)$
 & $-0.405162$ & $-0.11$ \\
 & $gg$ & $\phantom{-}8.957(87)$
 & $\phantom{-}9.048046$ & $-1.05$ \\
\midblackrule

\multirow{3}{*}{$1000$}
 & $q\bar q$ & $-3.8161(98)$
 & $-3.799312$ & $-1.71$ \\
 & $qg$ & $\phantom{-}6.742(12)$
 & $\phantom{-}6.732486$ & $+0.73$ \\
 & $gg$ & $\phantom{-}23.12(25)$
 & $\phantom{-}22.98039$ & $+0.55$ \\
\midblackrule

\multirow{3}{*}{$2000$}
 & $q\bar q$ & $\phantom{-}0.0169(86)$
 & $\phantom{-}0.007436$ & $+1.10$ \\
 & $qg$ & $\phantom{-}14.219(37)$
 & $\phantom{-}14.30716$ & $-2.35$ \\
 & $gg$ & $\phantom{-}60.75(38)$
 & $\phantom{-}61.20165$ & $-1.20$ \\
\bottomblackrule
\end{tabular}

\end{minipage}
\end{minipage}
\begin{minipage}{0.45\textwidth}
\begin{minipage}{1\textwidth}
{\fontsize{7.8}{8.8}\selectfont\bfseries
C - NLO corrections to total $pp\rightarrow t \bar{t}$ cross-section
\par}
\phantom{g}\\[-0.25cm]
\begin{tabular}{
@{\hspace{5pt}}
l@{\hspace{10pt}}
>{\columncolor{ludybg}[4pt][4pt]}l@{\hspace{12pt}}
>{\columncolor{mgfivebg}[4pt][4pt]}l@{\hspace{12pt}}
l@{\hspace{5pt}}
}
\topblackrule
\scalebox{0.9}{$\sqrt{s}$ [\scalebox{0.8}{$\text{GeV}$}]}
& \scalebox{0.9}{\texttt{LU} [\scalebox{0.8}{$\mathrm{pb}$}]}
& \scalebox{0.9}{\texttt{MG5aMC} [\scalebox{0.8}{$\mathrm{pb}$}]}
& \scalebox{0.9}{pull} \\
\midblackrule
$400$
& $1.0553(40)\!\cdot\!10^{-9}$
& $1.04980(71)\!\cdot\!10^{-9}$
& $+1.35$ \\

$1000$
& $4.407(30)\!\cdot\!10^{-3}$
& $4.4288(58)\!\cdot\!10^{-3}$
& $-0.71$ \\

$1500$
& $6.057(59)\!\cdot\!10^{-2}$
& $5.999(10)\!\cdot\!10^{-2}$
& $+0.97$ \\

$2000$
& $2.540(33)\!\cdot\!10^{-1}$
& $2.5239(50)\!\cdot\!10^{-1}$
& $+0.48$ \\
\bottomblackrule
\end{tabular}

\end{minipage}
\end{minipage}
\begin{minipage}{0.45\textwidth}
\begin{minipage}{1\textwidth}
{\fontsize{7.8}{8.8}\selectfont\bfseries
Parameters
\par}
\phantom{g}\\[-0.25cm]
\begingroup
\fontsize{7.4}{8.2}\selectfont
\begin{tabular}{
|@{\hspace{5pt}}
l@{\hspace{7pt}}!{\color{black!20!white}\vrule width 0.4pt}@{\hspace{7pt}}l
@{\hspace{7pt}}|@{\hspace{7pt}}
l@{\hspace{7pt}}!{\color{black!20!white}\vrule width 0.4pt}@{\hspace{7pt}}l
@{\hspace{5pt}}|
}
\hline
\rule{0pt}{3.1ex}$\alpha_s(m_Z^2)$ & $0.118$
& $1/\alpha$ & $132.507$ \\[0.5em]

$\mu_r=\Lambda$ & $91.188\,\mathrm{GeV}$
& $m_t$ & $173\,\mathrm{GeV}$ \\[0.3em]

$f_{\mathfrak q}^{\overline{\text{MS}}}$ &
\texttt{PDF4LHC21\_40} (ID $93100$)~\cite{PDF4LHCWorkingGroup:2022cjn}
& $e_q$ & $-e/3$ \\[3.1pt]
\hline
\end{tabular}
\endgroup

\end{minipage}
\end{minipage}

\caption{\justifying Results for the Drell--Yan and $t\bar{t}$-production processes, using Local Unitarity (\texttt{LU}), converted to the $\overline{\text{MS}}$ scheme. Results for \textbf{A} are obtained with 25M points for the full histogram and 2M refinements for the first three bins, with evaluation time at $\sim$40\,ms. Results for \textbf{B} are obtained with 100M points per channel and $\sqrt{\hat{s}}$, 400M for $gg$ at 400\,GeV, with typical evaluation times at $\sim$2--14\,ms, depending on the channel. Results for \textbf{C} are obtained with 200M points per $\sqrt{s}$, 100M for $\sqrt{s}=400$\,GeV, with typical evaluation times at $\sim$22\,ms. See \hyperref[sec:numerical_performance]{End Matter} for a discussion of numerical performance.}
{\color{gray!50}\rule{0.8\linewidth}{0.6pt}}
\label{fig:results}
\phantom{g}\\[-0.5cm]
\end{figure}

We exploit the IR finiteness of hadronic cross-sections in the $\text{LU}$ scheme to compute them in $d=4$ using Monte Carlo methods.
The finite hard shifts generated by $D^{-1}_{\mathfrak{q}\mathfrak{q}'}$ are added to the LU result to match the $\overline{\text{MS}}$ scheme.
For DY, one computes the bin contribution
\begin{equation}
\int_a^b \mathrm{d}z\frac{\mathrm{d}\sigma_{\mathfrak{q}\mathfrak{q}'}^{\text{LU},(1)}}{\mathrm{d}z}=\sum_{G\in\Gamma_{\mathfrak{q}\mathfrak{q}'}}\frac{\text{Re}\Delta_{ab}^G}{4\pi \langle\mathfrak{q}\rangle \langle\mathfrak{q}'\rangle},
\end{equation}
where
\begin{equation}
\Delta_{ab}^G=\! \int_a^b \mathrm{d}z\left[\prod_{i=1}^2\int_0^{\Lambda^2}\frac{\mathrm{d}p_i^2}{2\pi i}\text{disc}_{p_i^2}\right] \! N_{ee}I_{ G}^{\mathfrak{q}\mathfrak{q}'}.
\end{equation}
Writing discontinuities in terms of cut Feynman diagrams, and performing the residual (if any) energy integrations using the Loop Tree Duality~\cite{Catani:2008xa,Bierenbaum:2010cy,PhysRevLett.122.111603,Capatti:2019ypt} (or equivalent~\cite{JesusAguilera-Verdugo:2020fsn,Capatti:2020ytd,Capatti:2022mly,bobadilla2021lottylooptreeduality,Capatti:2023shz,Sborlini_2021}), we obtain the general form:
\begin{equation}
\Delta_{ab}^G=\! \int_{a}^b \mathrm{d}z\int\mathrm{d}\Pi_k N_{ee}\sum_{C_\sigma\in\mathcal{C}_G} \frac{\prod_{j\in C} \delta^{\sigma_j}(D_j)}{\prod_{i\in E\setminus C }D_i}N^{\mathfrak{q}\mathfrak{q}'}_G,
\end{equation}
where
\begin{equation}
\label{eq:measure_pik}
\mathrm{d}\Pi_k=\left[\prod_{i=1}^2\frac{\mathrm{d}p_i^2}{2\pi}\Theta(\Lambda^2-p_i^2)\right]\frac{\mathrm{d}^4 k}{(2\pi)^4},
\end{equation}
and $N^{\mathfrak{q}\mathfrak{q}'}_G$ is the numerator, a polynomial in $k, p_1^2, p_2^2$. $E$ is the set of propagators of $G$, with momenta $\ell_j$ and masses $m_j$, $D_j=\ell_j^2-m_j^2$, $\delta^{\pm}(D_j)\equiv\delta^{\pm}_{m_j^2}(\ell_j)$.
The sum runs over all tuples $C_\sigma=(\sigma,C)\in\mathcal{C}_G$ of cut sets $C\subset E$ and signs $\sigma\in\{\pm 1\}^{|C|}$, compatible with the energy integration and the expansion of the discontinuity in terms of cut diagrams.
Each contribution now has, for DY, $|C|=4$ (for $t\bar{t}$, $|C|=5$).
For a cut repeated propagator, the generalised cutting rule applies~\cite{Capatti:2022tit}.

The LP contribution is obtained by an asymptotic expansion in $\Lambda^2$, using the expansion-by-regions method~\cite{Chetyrkin:1988zz,Gorishnii:1989dd,etde_6475136,Beneke:1997zp,Jantzen:2011nz,Gardi:2022khw}. In principle, it is also possible to extract the leading power behaviour without a regions approach via a fitting procedure, similarly to what is done in slicing methods~\cite{Catani_2007}.
We find that all the regions are of hard, (anti-)collinear, soft or soft-(anti-)collinear~\footnote{We refer to ultrasoft regions as soft-(anti-)collinear since the transverse directions scale to zero faster than the longitudinal directions.} type:
\begin{equation}
\Delta_{ab}^G=\! \int_{a}^b \mathrm{d}z\int\mathrm{d}\Pi_k N_{ee}\sum_{C_\sigma, R} \frac{\prod_{j\in C} \delta^{\sigma_j}(D_j^R)}{\prod_{i\in E\setminus C }D_i^R}N^{\mathfrak{q}\mathfrak{q}'}_{G,R},
\end{equation}
where $D_i^R$ and $N^{\mathfrak{q}\mathfrak{q}'}_{G,R}$ are the propagators and numerator approximated in the region $R$.
The regions of the cut diagrams map to those of the uncut contribution $I_{ G}^{\mathfrak{q}\mathfrak{q}'}$.
The bare parton model is given by the hard region of $I_{ G}^{\mathfrak{q}\mathfrak{q}'}$.

We then introduce an auxiliary scaling variable $t$ in line with Soper's trick~\cite{Soper_2000,Capatti:2020xjc}, rescaling $z\rightarrow t^2 z$ and $\mathbf{k}\rightarrow t \mathbf{k}$, and solve the four Dirac delta functions in $p_1^2,p_2^2,k^0,t$.
That gives an integrand that is locally free of infrared singularities and can be directly fed to a Monte Carlo integrator.
Finally, ultraviolet singularities are renormalised locally by the BPHZ-based~\cite{BogoliubowParasiuk1957,Zimmermann1969,Hepp1966} procedure of Ref.~\cite{Capatti:2022tit}, introducing dependence on a renormalisation scale $\mu_r$, which can be varied independently from the factorisation scale $\Lambda$.
Thresholds of loop diagrams are subtracted using the procedure of Ref.~\cite{Kermanschah:2021wbk}, being careful to correctly zero their contribution close to soft singularities (see Fig.~\ref{fig:triangle_region}).

We report in Fig.~\ref{fig:results} Monte Carlo results for the DY (differential in $z$) and $pp\rightarrow t \bar{t}$ (inclusive) hadronic cross-sections at NLO, using a path-finder code, which uses \href{https://symbolica.io/}{\texttt{Symbolica}} and several libraries developed in the context of \href{https://github.com/alphal00p/gammaloop}{\texttt{gammaloop}}.
We use one $d$-quark flavour, $q=d$.
Agreement with the $\overline{\text{MS}}$ partonic cross-sections~\cite{Altarelli:1979ub,Czakon:2008ii, 1988NuPhB.303..607N} for all channels is achieved after the scheme conversion Eq.~\eqref{eq:splitting_cts_nlo} (with $b_\partial^{\mathfrak{q}\mathfrak{q}'}=0$), confirming that $D_{gg}$ and $D_{gq}$ are also correct.
The hadronic result, compared against $\texttt{MG5aMC v3.7.0}$~\cite{Alwall:2014hca}, required the decoupling-scheme counterterm to obtain agreement.

\section{Outlook}

We successfully constructed hadronic cross-sections in terms of partonic cross-sections with an arbitrary number of initial-state partons, which are infrared finite %
and compatible with collinear factorisation.
We tested the approach through analytic and Monte Carlo computations of three QCD cross-sections, developing in the process a \emph{fully local} approach to the calculation.
Two possible future research directions are:

\begin{itemize}
\item[\textit{i}] \textit{Generalisation to NNLO LP}: extending the formalism to NNLO would allow us to compute hadronic cross-sections entirely by Monte Carlo methods, including loop integrals, in momentum space.
This programme is core to both the Local Unitarity and the ETHZ--Stony Brook subtraction frameworks~\cite{Anastasiou:2018rib,Anastasiou:2020sdt,Anastasiou:2022eym,Anastasiou:2024xvk,Anastasiou:2026kpm,Capatti:2019ypt,Capatti:2019edf,Kermanschah:2021wbk,Kermanschah:2024utt,Kermanschah:2025wlo}, and has the potential to overcome current computational bottlenecks in the production of precise fixed-order predictions.

\item[\textit{ii}] \textit{Power corrections}: The $\Lambda^2$ expansion in the formalism is controlled, and the subleading-power cross-section can be written in terms of cut Feynman integrals. KLN-based approaches have been used to probe linear power corrections~\cite{Akhoury_1996,Akhoury:1995sp,Fontes:2026uji}.
A natural step is to compute subleading-power corrections for DIS, where the Operator Product Expansion is understood at subleading twist~\cite{JAFFE1981467,Ellis:1982cd,Blumlein:2008kz}.
Beyond DIS, comparisons with operator~\cite{Qiu:1990xxa,Feige2017SubleadingFactorization,Moult:2018jjd,Moult2017SubleadingBasis,Moult2019RadiativeFunctions,Beneke2018SubleadingNJet,Beneke2019NLPDrellYan,Beneke2020ThresholdFactorization}, dispersive~\cite{Dokshitzer1998Milan,DokshitzerMarchesiniWebber1996Dispersive,DokshitzerWebber1995PowerCorrections} and massified-gluon/renormalon~\cite{Korchemsky_1995,NasonSeymour1995Renormalons,BenekeBraun1995DrellYan,FerrarioRavasio2019Renormalons,Caola:2021kzt,Caola:2022vea,Fontes:2026uji} approaches can be considered.

\end{itemize}

Finally, the infrared interplay between the multi-partonic contributions and the parton-model diagrams could be a basis for studies of factorisation and factor\-isation-breaking~\cite{Collins:2007nk,Rogers:2010dm,Catani:2011st,Forshaw:2012bi,Gaunt:2014ska,Zeng:2015iba,Rothstein:2016bsq,Schwartz:2017nmr,Becher:2023lmy,Becher:2024vdk}, especially due to our inclusion of spectator particles. Our framework is uniquely positioned to simultaneously attack these questions.
\\[0.3em]

\begin{acknowledgments}
\paragraph{Acknowledgements} We are grateful to Valentin Hirschi for his invaluable support and input, and to Michael Ruf for several discussions.
This work was supported in part by the Swiss National Science Foundation (SNSF) under grant No. PCEFP2\_203335. The work of Z.C. is supported by the European Union's Horizon Europe research and innovation programme under the Marie Skłodowska-Curie Actions (grant agreement No. 101271880). The views expressed are solely those of the authors and do not necessarily reflect those of the European Union or the European Research Executive Agency (REA), neither of which accepts responsibility for them.
\textbf{LLM use}: the authors hold ChatGPT Pro subscriptions and used the GPT-5 and GPT-6 models during the development of this work. All analytical calculations reported in the paper were performed at least once by a human without AI assistance. The path-finder code was developed with substantial AI assistance, particularly after the release of GPT-5.4. The manuscript was written by the authors; AI tools were only used in an editorial capacity.

\end{acknowledgments}

\bibliography{apssamp}

\section{End Matter}
\subsection{List of integrals for DIS and DY}
\label{sec:integrals_DIS_DY}

\begin{figure}[t]
\includegraphics[width=\linewidth]{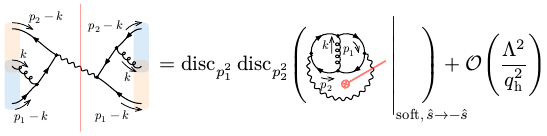}
\caption{\justifying An example of a non-trivial identification, at leading power, of cut diagrams with discontinuities. In this case, the interference diagram where the gluon belongs to both jets at the same time is equated to the double discontinuity of the soft contribution to the integral $I_{\triangle}^{q\bar{q}}$, represented as an embedding, with the continuation $\hat{s}\to-\hat{s}$.}
\label{fig:triangle_soft}
{\color{gray!50}\rule{0.8\linewidth}{0.6pt}}
\phantom{g}\\[-0.3cm]
\end{figure}

We provide a list of the integrals appearing in Eq.~\eqref{eq:disc_dis} and Eq.~\eqref{eq:disc_dy}.
At LP, their discontinuities can be computed either by computing the integrals and then using the usual definition of discontinuity, or by writing the discontinuity as a sum over cut Feynman integrals (corresponding to the original interference diagrams). The equivalence between the two definitions is non-trivial, but can be shown using the expansion-by-regions method (see, e.g., Fig.~\ref{fig:triangle_soft}) or by direct computation.

\textbf{DIS:} For DIS, we have $\Gamma_{q}=\{\triangle,\Box,\bigcirc,\text{lsz}\}$ and $\Gamma_{g}=\{\Box,\timesbox\}$.
The corresponding integrals have the form
\begin{equation}
\frac{I^{\mu\nu}_{\mathfrak{q}G}}{8\pi^2 e_q^2}= \bar{\mu}^{2\epsilon}C_FN_c \int \frac{\mathrm{d}^d k}{(2\pi)^d} \frac{N_{\mathfrak{q}G}^{\mu\nu}}{\prod_{j=1}^6 D_j^{\alpha_j^G}},
\end{equation}
with $D_1=k^2$, $D_2=(k+p)^2$, $D_3=(k+p+q)^2$, $D_4=(k+q)^2$, $D_5=p^2$, $D_6=(p+q)^2$.
The factor $\bar{\mu}^{2\epsilon}$ is the usual $\overline{\text{MS}}$ measure, and $\boldsymbol{\alpha}^\triangle=[1,1,1,0,1,1]$, $\boldsymbol{\alpha}^\Box=[1,2,1,0,1,0]$, $\boldsymbol{\alpha}^{\timesbox}=[1,1,1,1,1,0]$, $\boldsymbol{\alpha}^{\bigcirc}=[1,0,1,0,1,2]$, $\boldsymbol{\alpha}^{\text{lsz}}=[1,1,0,0,2,1]$.
The numerators including graph factors are
\begin{align}
\hspace{-0.1cm}N_{q\triangle}^{\mu\nu}&=-2\text{Tr}[\slashed{p}\gamma^\sigma (\slashed{k}+\slashed{p}) \gamma^\mu (\slashed{k}+\slashed{p}+\slashed{q})\gamma_\sigma (\slashed{p}+\slashed{q})\gamma^\nu]\\
\hspace{-0.1cm}N_{q\Box}^{\mu\nu}&=-\text{Tr}[\slashed{p}\gamma^\sigma (\slashed{k}+\slashed{p}) \gamma^\mu (\slashed{k}+\slashed{p}+\slashed{q})\gamma^\nu (\slashed{k}+\slashed{p})\gamma_\sigma]\\
\hspace{-0.1cm}N_{q\bigcirc}^{\mu\nu}&=-\text{Tr}[\slashed{p}\gamma^\mu (\slashed{p}+\slashed{q}) \gamma^\sigma (\slashed{k}+\slashed{p}+\slashed{q})\gamma_\sigma (\slashed{p}+\slashed{q})\gamma^\nu] \\
\hspace{-0.1cm}N_{q\text{lsz}}^{\mu\nu}&=-\text{Tr}[\slashed{p}\gamma^\mu (\slashed{p}+\slashed{q}) \gamma^\nu \slashed{p} \gamma^\sigma (\slashed{k}+\slashed{p})\gamma_\sigma]\\
\hspace{-0.1cm}N_{g\Box}^{\mu\nu}&=2\text{Tr}[\slashed{k}\gamma^\sigma (\slashed{k}+\slashed{p}) \gamma^\mu (\slashed{k}+\slashed{p}+\slashed{q})\gamma^\nu (\slashed{k}+\slashed{p})\gamma_\sigma]\\
\hspace{-0.1cm}N_{g\timesbox}^{\mu\nu}&=2\text{Tr}[\slashed{k}\gamma^\sigma (\slashed{k}+\slashed{p}) \gamma^\mu (\slashed{k}+\slashed{p}+\slashed{q})\gamma_\sigma (\slashed{k}+\slashed{q})\gamma^\nu].
\end{align}

\textbf{DY, zero bin:} For DY, we have $\Gamma_{q\bar{q}}=\{\triangle,\allowbreak \text{lsz}_1,\allowbreak\text{lsz}_2,\allowbreak\Box_1,\allowbreak\Box_2,\allowbreak \timesbox\}$ and $\Gamma_{gq}=\{\slashed{\triangle},\allowbreak\bigcirc,\allowbreak \Box\}$.
The $q\bar{q}$ channel contributions that are localised at $\hat{s}=q^2$ read:
\begin{equation}
\begin{aligned}
\frac{I_{G}^{q\bar{q}}}{8\pi^2 e_q^2 }&=-\bar{\mu}^{2\epsilon}C_FN_c \delta_{q^2}^+(p_{12})\Bigg[
\frac{\hat{s}\, \delta_{G\triangle}}
{2p_1^2p_2^2}+\hspace{-0.1cm}\int \hspace{-0.1cm}\frac{\mathrm{d}^d k}{(2\pi)^d i} \frac{N_{G}^{q\bar{q}} }{\prod_{j=1}^5 D_j^{\alpha_j^G}}\Bigg],
\end{aligned}
\end{equation}
for $G\in\{\triangle,\text{lsz}_1,\text{lsz}_2\}$, with $D_1=k^2$, $D_2=(k+p_1)^2$, $D_3=(k-p_2)^2$, $D_4=p_1^2$, $D_5=p_2^2$, $p_{12}=p_1+p_2$. The first term in the bracket arises by matching the soft region of the triangle to its respective interference diagram arising from the KLN sum, which amounts to the mapping $\hat{s}\rightarrow -\hat{s}$ (see Fig.~\ref{fig:triangle_soft}).
We have $\boldsymbol{\alpha}^{\triangle}=[1,1,1,1,1]$, $\boldsymbol{\alpha}^{\text{lsz}_1}=[1,1,0,2,1]$, $\boldsymbol{\alpha}^{\text{lsz}_2}=[1,0,1,1,2]$, and
\begin{align}
N_{\triangle}^{q\bar{q}}&=2\text{Tr}[\slashed{p}_1\gamma^\sigma (\slashed{k}+\slashed{p}_1) \gamma^\mu (\slashed{k}-\slashed{p}_2)\gamma_\sigma \slashed{p}_2\gamma_\mu], \\
N_{\text{lsz}_1}^{q\bar{q}}&=\text{Tr}[\slashed{p}_1\gamma^\sigma (\slashed{k}+\slashed{p}_1) \gamma_\sigma \slashed{p}_1\gamma_\mu \slashed{p}_2\gamma^\mu],
\end{align}
and $N_{\text{lsz}_2}^{q\bar{q}}=N_{\text{lsz}_1}^{q\bar{q}}|_{p_1\leftrightarrow p_2,k\rightarrow -k}$.
The factor of $2$ for $N_{\triangle}^{q\bar{q}}$ accounts for the complex conjugate graph.

\textbf{DY, spectrum:} For the diagrams that contribute to the entire $z=q^2/\hat{s}$ spectrum, namely $G\in\{\slashed{\triangle},\allowbreak\Box,\allowbreak\Box_1,\allowbreak\Box_2,\allowbreak \timesbox,\allowbreak\bigcirc\}$, we have
\begin{equation}
\frac{I_{G}^{\mathfrak{q}\mathfrak{q}'}}{8\pi^2e_q^2 }=\bar{\mu}^{2\epsilon}C_FN_c\int \frac{\mathrm{d}^d k}{(2\pi)^d} \frac{N_{G}^{\mathfrak{q}\mathfrak{q}'} \, \delta_{q^2}^+(p_{12}-k)\delta^+_0(k)}{\prod_{j=1}^5 D_j^{\alpha_j^G}},
\end{equation}
with $D_1=(k-p_1)^2$, $D_2=(k-p_2)^2$, $D_3=p_1^2$, $D_4=p_2^2$ and $D_5=p_{12}^2$.
We now have $\boldsymbol{\alpha}^{\slashed{\triangle}}=[1,0,1,1,1]$, $\boldsymbol{\alpha}^{\Box}=\boldsymbol{\alpha}^{\Box_1}=[2,0,1,1,0]$, $\boldsymbol{\alpha}^{\Box_2}=[0,2,1,1,0]$, $\boldsymbol{\alpha}^{\timesbox}=[1,1,1,1,0]$, $\boldsymbol{\alpha}^{\bigcirc}=[0,0,1,1,2]$.
The numerators read:
\begin{align}
N^{gq}_{\slashed{\triangle}}&=2\text{Tr}[\slashed{p}_2\gamma^\sigma (\slashed{k}-\slashed{p}_1) \gamma^\mu \slashed{k}\gamma_\sigma (\slashed{p}_1+\slashed{p}_2)\gamma_\mu], \\
N^{gq}_{\Box}&=\text{Tr}[\slashed{p}_2\gamma^\sigma (\slashed{k}-\slashed{p}_1) \gamma_\mu \slashed{k}\gamma^\mu (\slashed{k}-\slashed{p}_1)\gamma_\sigma],\\
N^{gq}_{\bigcirc}&=\text{Tr}[\slashed{p}_2\gamma^\sigma \slashed{p}_{12} \gamma_\mu \slashed{k}\gamma^\mu \slashed{p}_{12}\gamma_\sigma],
\\
N^{q\bar{q}}_{\Box_1}&=\text{Tr}[\slashed{p}_1\gamma^\sigma (\slashed{k}-\slashed{p}_1) \gamma_\mu \slashed{p}_2\gamma^\mu (\slashed{k}-\slashed{p}_1)\gamma_\sigma],\\
N^{q\bar{q}}_{\timesbox}&=-2\text{Tr}[\slashed{p}_1\gamma^\sigma (\slashed{k}-\slashed{p}_1) \gamma^\mu \slashed{p}_2\gamma_\sigma (\slashed{k}-\slashed{p}_2)\gamma_\mu],
\end{align}
and $N_{\Box_2}^{q\bar{q}}=N_{\Box_1}^{q\bar{q}}|_{p_1\leftrightarrow p_2}$.

\subsection{Reproducible analytic calculations}
\label{sec:reproducible-analytic-calculations}

The \href{https://github.com/ZenoCapatti/LUDY-resources}{LUDY-resources repository} hosts Python notebooks for the analytic calculations of the NLO DIS and DY cross-sections and the splitting counterterms. It also provides extensive information on installation and usage.

\textbf{DIS and DY cross-sections}: the DIS and DY cross-sections are computed by generating the routed graphs and their cuts, performing the Dirac and colour algebra and reducing the scalar integrals to master integrals. In line with Reverse Unitarity, only the master integrals possessing the same cuts as the original diagram contribute. The master integrals themselves are simple and are evaluated by direct integration. The integrated results are then expanded in virtuality, to give the LP result. The computation of the parton-model diagrams is benchmarked against the DIS coefficient functions in Chapter~4 of Ref.~\cite{Ellis:1996mzs} and the Drell--Yan cross-sections of Ref.~\cite{Altarelli:1979ub}.

\textbf{Splitting counterterms}: a separate notebook derives the splitting counterterms $D_{qq}$, $D_{gq}$, $D_{qg}$, and $D_{gg}$ in the axial gauge and for arbitrary NLO processes. In the axial gauge, collinear physics is entirely contained in self-energy corrections to external legs or squared t-channel exchanges between a parton and a hard sub-process, independently of the specific initial-state cuts. We start from these singular diagrams, perform the kinematic and colour algebra, expand in $\Lambda^2$ at the integrand level (using the expansion-by-regions method), and perform the $k$ and $p_i^2$ integrations. The Dirac delta functions fix the value of $p_i^2$ and $k_{-}$. When a raised propagator is cut, a derivative in $p_i^2$ has to be taken: the integral corresponding to the derivative hitting on the upper boundary $\Lambda^2$ gives the boundary term $b_{\partial}^{\mathfrak{q}\mathfrak{q}'}$.

\textbf{Tools}: graphs are generated with \texttt{Feyngen}~\cite{GammaLoopSoftware}, while numerator algebra is performed with \texttt{Spenso} and \texttt{Idenso}~\cite{SpensoSoftware}. All symbolic manipulations use \texttt{Symbolica}~\cite{SymbolicaSoftware}. IBP reduction is obtained with a custom Python API for \texttt{Kira}~\cite{Klappert_2021}.

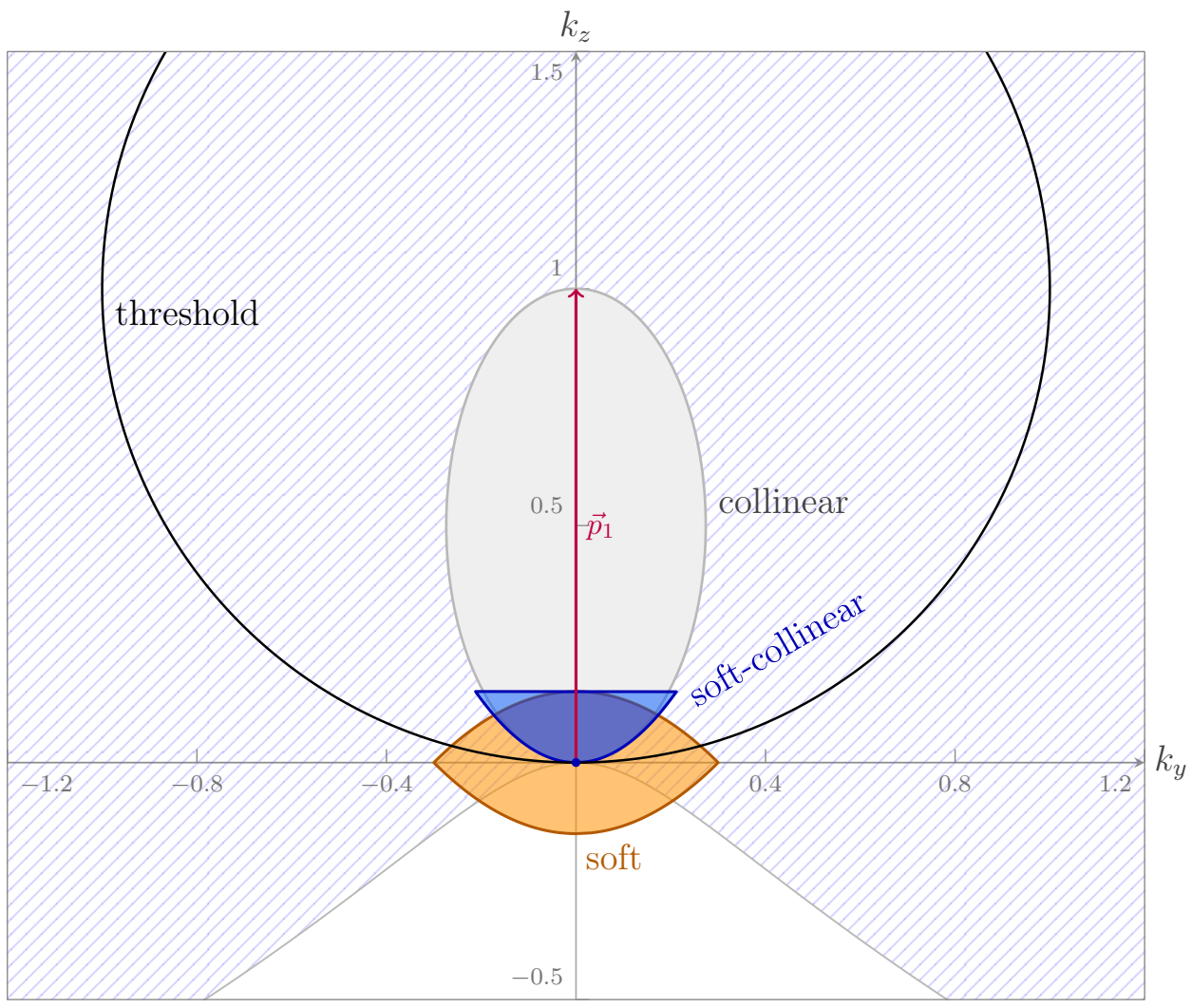
\begin{figure}[!tp]
\scalebox{0.6}{
\begin{tikzpicture}[line join=round,line cap=round]
\begin{axis}[
  width=13.2cm, height=11cm,
  scale only axis,
  axis equal image,
  xmin=-1.2, xmax=1.2,
  ymin=-0.5, ymax=1.5,
  xtick={-1.2,-0.8,-0.4,0.4,0.8,1.2},
  ytick={-0.5,0.5,1,1.5},
  xticklabels={,,,,,}, yticklabels={,,,},
  tick label style={black!55,font=\footnotesize},
  tick style={black!45,line width=0.45pt},
  axis lines=middle,
  axis line style={black!45,line width=0.55pt},
  xlabel={$k_y$}, ylabel={$k_z$},
  xlabel style={anchor=west,text=black!75,font=\large},
  ylabel style={anchor=south,text=black!75,font=\large},
  tick align=inside,
  clip=false,
]

\begin{scope}
\clip (axis cs:-1.2,-0.5) rectangle (axis cs:1.2,1.5);

\path[
  fill=red,
  fill opacity=0.26,
  pattern=north east lines,
  pattern color=blue!60
]
  (axis cs:-2,1) rectangle (axis cs:2,2);

\path[
  fill=red,
  fill opacity=0.26,
  pattern=north east lines,
  pattern color=blue!60
]
  plot[domain=0:1,samples=220,variable=\z]
    (axis cs:{sqrt(max(0,0.3*\z*(1-\z)))},{\z})
  -- (axis cs:2,1) -- (axis cs:2,0) -- cycle;

\path[
  fill=red,
  fill opacity=0.26,
  pattern=north east lines,
  pattern color=blue!60
]
  plot[domain=1:0,samples=220,variable=\z]
    (axis cs:{-sqrt(max(0,0.3*\z*(1-\z)))},{\z})
  -- (axis cs:-2,0) -- (axis cs:-2,1) -- cycle;

\path[
  fill=red,
  fill opacity=0.26,
  pattern=north east lines,
  pattern color=blue!60
]
  plot[domain=-1:0,samples=260,variable=\z]
    (axis cs:{sqrt(max(0,-0.3*(\z-1)^2*\z/(0.7+0.3*\z)))},{\z})
  -- (axis cs:2,0) -- (axis cs:2,-1) -- cycle;

\path[
  fill=red,
  fill opacity=0.26,
  pattern=north east lines,
  pattern color=blue!60
]
  plot[domain=0:-1,samples=260,variable=\z]
    (axis cs:{-sqrt(max(0,-0.3*(\z-1)^2*\z/(0.7+0.3*\z)))},{\z})
  -- (axis cs:-2,-1) -- (axis cs:-2,0) -- cycle;

\path[
  fill=white,
  fill opacity=0.28,
  draw=gray!55,
  line width=0.6pt
]
  (axis cs:0,0)
  -- plot[domain=0:-1,samples=260,variable=\z]
    (axis cs:{sqrt(max(0,-0.3*(\z-1)^2*\z/(0.7+0.3*\z)))},{\z})
  -- plot[domain=-1:0,samples=260,variable=\z]
    (axis cs:{-sqrt(max(0,-0.3*(\z-1)^2*\z/(0.7+0.3*\z)))},{\z})
  -- cycle;

\path[
  fill=black!30,
  fill opacity=0.20,
  draw=gray!55,
  line width=0.85pt
]
  plot[domain=0:360,samples=240,variable=\t]
    (axis cs:{sqrt(0.075)*cos(\t)},{0.5+0.5*sin(\t)})
  -- cycle;

\path[
  fill=orange!85!yellow,
  fill opacity=0.55,
  draw=orange!70!black,
  line width=0.95pt
]
  plot[domain=-0.15:0,samples=100,variable=\z]
    (axis cs:{sqrt(max(0,0.6*(\z+0.15)))},{\z})
  -- plot[domain=0:0.15,samples=100,variable=\z]
    (axis cs:{sqrt(max(0,0.6*(0.15-\z)))},{\z})
  -- plot[domain=0.15:0,samples=100,variable=\z]
    (axis cs:{-sqrt(max(0,0.6*(0.15-\z)))},{\z})
  -- plot[domain=0:-0.15,samples=100,variable=\z]
    (axis cs:{-sqrt(max(0,0.6*(\z+0.15)))},{\z})
  -- cycle;

\path[
  fill=blue!65!cyan,
  fill opacity=0.50,
  draw=blue!70!black,
  line width=0.95pt
]
  plot[domain=0:0.15,samples=100,variable=\z]
    (axis cs:{sqrt(max(0,0.3*\z))},{\z})
  -- plot[domain=0.15:0,samples=100,variable=\z]
    (axis cs:{-sqrt(max(0,0.3*\z))},{\z})
  -- cycle;

\path[fill=blue,fill opacity=0.20]
  plot[domain=0:0.1,samples=100,variable=\z]
    (axis cs:{sqrt(max(0,0.3*\z))},{\z})
  -- plot[domain=0.1:0.15,samples=100,variable=\z]
    (axis cs:{sqrt(max(0,0.09-0.6*\z))},{\z})
  -- plot[domain=0.15:0.1,samples=100,variable=\z]
    (axis cs:{-sqrt(max(0,0.09-0.6*\z))},{\z})
  -- plot[domain=0.1:0,samples=100,variable=\z]
    (axis cs:{-sqrt(max(0,0.3*\z))},{\z})
  -- cycle;

\draw[black,thick]
  plot[domain=0:360,samples=360,variable=\t]
    (axis cs:{cos(\t)},{1+sin(\t)})
  -- cycle;
\end{scope}

\draw[gray,line width=0.4pt]
  (axis cs:-1.2,-0.5) rectangle (axis cs:1.2,1.5);

\begin{scope}[every node/.style={text=black!55,font=\footnotesize,inner sep=1pt}]
  \pgfplotsinvokeforeach{-0.8,-0.4,0.4,0.8}{
    \node[anchor=north,yshift=-3pt] at (axis cs:#1,0) {$#1$};
  }
  \node[anchor=north west,xshift=3pt,yshift=-3pt] at (axis cs:-1.2,0) {$-1.2$};
  \node[anchor=north east,xshift=-3pt,yshift=-3pt] at (axis cs:1.2,0) {$1.2$};
  \pgfplotsinvokeforeach{0.5,1}{
    \node[anchor=south east,xshift=-3pt,yshift=3pt] at (axis cs:0,#1) {$#1$};
  }
  \node[anchor=north east,xshift=-3pt,yshift=-3pt] at (axis cs:0,1.5) {$1.5$};
  \node[anchor=south east,xshift=-3pt,yshift=3pt] at (axis cs:0,-0.5) {$-0.5$};
\end{scope}

\addplot[red!75!blue,line width=1.0pt,->]
  coordinates {(0,0) (0,1)}
  node[midway,right,text=red!75!blue,font=\normalsize] {$\vec{p}_1$};

\node[below right,text=orange!70!black,font=\large]
  at (axis cs:0,-0.15) {soft};

\node[
  below right,
  text=blue!70!black,
  font=\large,
  rotate=30
] at (axis cs:0.2,0.17) {soft-collinear};

\node[below right,text=black!75,font=\large]
  at (axis cs:0.28,0.6) {collinear};

\node[text=black,font=\large]
  at (axis cs:-0.82,0.95) {threshold};

\addplot[
  only marks,
  mark=*,
  mark size=1.25pt,
  blue!70!black
] coordinates {(0,0)};

\end{axis}
\end{tikzpicture}
}
\caption{\justifying For the infrared-finite class of the DY process containing the virtual triangle diagram, the cancelling bulk singularities are located at the soft origin point, on the (anti-)collinear segment (in red), and on the black threshold circle. The parton-model diagram contributes with a hard region, and is non-zero everywhere in the loop momentum ($k$) space. Each additional multi-partonic cut has support only in compact regions (the shaded ``collinear'', ``soft-collinear'', and ``soft'' regions), due to the theta functions in Eq.~\eqref{eq:measure_pik}, while the threshold counterterm has support in the striped region, designed to exclude the collinear segment and so that the principal value of the counterterm is zero.
In the tiny sliver obtained by subtracting the collinear region from the soft-collinear region, the integrand is unbounded. The sliver narrows toward the soft point, yielding integrable behaviour.}
{\color{gray!50}\rule{0.8\linewidth}{0.6pt}}
\phantom{g}\\[-0.3cm]
\label{fig:triangle_region}
\end{figure}

\subsection{Numerical performance}
\label{sec:numerical_performance}

Our path-finder Local Unitarity implementation provides a blueprint for future high-performance implementations, including within the \href{https://github.com/alphal00p/gammaloop}{\texttt{gammaloop}} framework.
As expected, it underperforms established tools such as \texttt{MG5aMC}, allowing us to clearly identify two performance issues to be addressed in future work.

First, at $\sqrt{s}=2\,\mathrm{TeV}$, typical integrand evaluation times for $pp\rightarrow t\bar{t}$ are approximately $22\,\mathrm{ms}$ per sample and core on a dual-socket AMD EPYC 9754 compute node, using double precision with rotational stability checks and retries at 32- and 60-digit precision when necessary.
This is consistent with a non-optimised workflow that uses efficient Symbolica evaluators but may be limited by inefficient symbolic manipulations and Python orchestration.
Based on our experience, using a low-level language such as Rust and carefully managing the size of the symbolic expressions subject to evaluation could lower evaluation times by at least an order of magnitude.

The second is slow convergence, which appears largely attributable to two aspects of the diagrams' soft behaviour:
\begin{itemize}
\item The variance of the integrand for the hadronic $pp\rightarrow t\bar{t}$ cross-sections increases with $\sqrt{s}$, consistent with the cancellation of double soft logarithms across graphs ($\sim$0 digits at 400\,GeV and $\sim$1 digit at 2\,TeV).
A solution would be to realise the cancellation of these soft logarithms locally, by carefully aligning the phase spaces of virtual and real emission diagrams.

\item The soft region, especially for infrared-finite classes containing at least one virtual diagram, involves delicate cancellations between cut diagrams and threshold counterterms (see Fig.~\ref{fig:triangle_region}). Larger weights arise in increasingly narrow regions around the soft point, yielding integrable behaviour but slowing convergence. Future high-performance implementations would need to address this through an appropriate parametrisation around the soft point.
\end{itemize}

Since our implementation aimed primarily to demonstrate the correctness of the theoretical ideas presented here and lay the foundations for the Local Unitarity approach, we leave these performance issues to future work.

\phantom{g}\\
\phantom{g}\\
\phantom{g}

\end{document}